\documentclass{svjour3} 
\smartqed  % flush right qed marks, e.g. at end of proof
\usepackage{graphicx,amsmath,enumerate}

\usepackage{bm,hyphenat,xspace}
\usepackage{graphicx,epsfig}

\newcommand {\mcu}{\mathcal{U}}

\begin{document}

\title {Dimer-atom-atom recombination in the universal four-boson system
\footnote{Dedicated to Professor Henryk Witala at the
 occasion of his 60th birthday}}

\author{A. Deltuva} 
\institute{ A. Deltuva  \at
Centro de F\'{\i}sica Nuclear da Universidade de Lisboa,
P-1649-003 Lisboa, Portugal \\
              \email{deltuva@cii.fc.ul.pt} } 

\date{Received: date / Accepted: date}

 \maketitle

\begin{abstract}
The dimer-atom-atom recombination process in the
system of four identical bosons with resonant interactions
is studied. The description uses the 
exact Alt, Grassberger and Sandhas equations for the four-particle 
transition operators that are solved in the momentum-space framework.
The dimer-dimer and atom-trimer channel
contributions to the ultracold dimer-atom-atom recombination rate
 are calculated. The dimer-atom-atom recombination rate greatly exceeds
the three-atom recombination rate.
\end{abstract}

%%%%%%%%%%%%%%%%%%%%%%%%%%%%%%%%%%%%%%%%%%%%%%%%%%%%%%%%%%%%%%%%%%%%%%%%%%%%%%%
\section{Introduction}

Few-particle systems with large two-particle 
scattering length $a$ possess universal properties that
are independent of the short-range interaction details.
The three-particle system was investigated theoretically
by V. Efimov more than 40 years ago \cite{efimov:plb}
but only in the last decade the cold-atom physics
experiments \cite{kraemer:06a} confirmed
his prediction for the existence of zero orbital angular momentum weakly
bound three-particle states  with asymptotic discrete scaling symmetry.
%Theoretical and experimental studies were
This boosted the interest also in the universal systems 
with four or even more particles. In contrast to
the three-body system where semi-analytical
results have been obtained (see Ref.~\cite{braaten:rev} for a review),
most of the theoretical studies of the four-body systems 
are numerical. In addition to the numerous bound-state calculations, e.g., 
\cite{blume:00a,hammer:07a,stecher:09a,yamashita:06a,lazauskas:he,yamashita:10a,schmidt:10a,gattobigio:12a},
also the collision processes have been investigated
in the framework of hyperspherical harmonics (HH) \cite{stecher:09a},
coordinate-space Faddeev-Yakubovsky (FY) equations 
\cite{yakubovsky:67,lazauskas:he}
or momentum-space Alt,Grassberger and Sandhas (AGS) equations
\cite{grassberger:67,deltuva:10c}.
The studies include the elastic and inelastic atom-trimer  
\cite{lazauskas:he,deltuva:10c,deltuva:11a}
and dimer-dimer scattering \cite{dincao:09a,deltuva:11b}
as well as the four-atom recombination 
\cite{PhysRevLett.102.133201,stecher:09a,mehta:09a,deltuva:12a}.
Depending on the collision energy and the two-particle
scattering length (that can be manipulated by the external
magnetic field in the cold-atom experiments)
all these reactions may get resonantly enhanced due to
the presence of unstable tetramer states.
Such a resonant behaviour in the dimer-dimer relaxation
and four-atom recombination, roughly consistent with theoretical predictions,
 was observed recently in experiments with ultracold atoms 
performed near the universal regime
\cite{ferlaino:08a,ferlaino:09a,pollack:09a,zaccanti:09a}.

The four-particle reaction, that, to the best of our knowledge,
has not been studied so far, is the three-cluster recombination, i.e., 
an inelastic collision of a dimer and two atoms leading to the
two-cluster final state either with two dimers or with one atom and a trimer.
Due to time reversal symmetry, the amplitude for this process
is equal to the  amplitude for the three-cluster
breakup of the initial two-cluster state. 
The latter is more directly related to the transition operators
for the two-cluster reactions 
\cite{deltuva:10c,deltuva:11b,deltuva:12a} and is derived
in the present work using the formalism of the AGS 
four-particle scattering equations. Their numerical
solution is performed in the momentum-space framework.

\begin{figure}[!]
\includegraphics[scale=0.5]{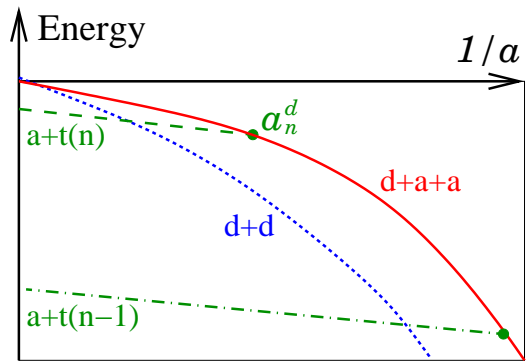}
\caption{\label{fig:e-a} (Color online)
Schematic representation of the two- and three-cluster
thresholds in the four-boson system as functions 
of the two-boson scattering length.
The intersection of the dimer-atom-atom (solid line)
and the $n$th atom-trimer (dashed line) thresholds
defines  $a_n^d$. }
\end{figure}

We will study the dimer-atom-atom recombination 
in the system of four identical bosons with resonant two-boson
interactions. In the potential model taken over from  Refs.
\cite{deltuva:10c,deltuva:11b,deltuva:12a} there is only
one weakly bound dimer but many trimers since the system exhibits 
an Efimov effect \cite{efimov:plb,braaten:rev}. 
In our nomenclature  we label the Efimov trimers by one integer number
$n$, starting with $n=0$ for the ground state.
The dependence of various scattering thresholds on the two-boson
scattering length $a$ is schematically shown in 
Fig.~\ref{fig:e-a}.  The special value of $a$ corresponding
to the intersection of the $n$-th atom-trimer and 
dimer-atom-atom threshold is denoted by $a_n^d$.

In Sec.~\ref{sec:4bse} we  derive the  three-cluster
breakup amplitude in terms of the AGS transition operators
and discuss some technical aspects of the calculations.
In Sec.~\ref{sec:res} we present results for the 
dimer-atom-atom recombination.
We summarize in  Sec.~\ref{sec:sum}.

\section{Four-particle scattering \label{sec:4bse}}

We consider the relative motion in the four-particle system
interacting via short-range pairwise potentials $v_j$ 
where $j$ labels one of the six pairs.
The full Hamiltonian is
\begin{equation} \label{eq:H}
H = H_0 + \sum_{j=1}^6 v_j 
\end{equation}
with  $H_0$ being the kinetic energy operator.
We do not calculate the four-particle
wave function directly by solving the Schr\"odinger equation
or the Faddeev-Yakubovsky equations \cite{yakubovsky:67}.
We prefer the integral form of the scattering equations as formulated
by Alt, Grassberger and Sandhas  \cite{grassberger:67}
for the  four-particle transition operators $\mcu_{\sigma \rho}^{ji}$.
These  equations are 18-component
matrix equations \cite{grassberger:67,fonseca:87} , i.e.,
\begin{equation} \label{eq:AGSgen}
%\begin{split}
\mcu_{\sigma \rho}^{ji} = %{} & 
(G_0\, t_i\, G_0)^{-1}\,
\bar{\delta}_{\sigma \rho} \, \delta_{ji} 
  + \sum_{\gamma k} \bar{\delta}_{\sigma \gamma}  U_{\gamma}^{jk} G_0\, t_k\, 
G_0 \, \mcu_{\gamma \rho}^{ki},
%\end{split}
\end{equation}
where the components are distinguished by the chains of partitions, i.e.,
by the two-cluster partition 
and by the three-cluster partition. The two-cluster partitions,
denoted by Greek letters, are either of $3+1$ or $2+2$ type.
The  three-cluster partitions, denoted by Latin letters,
 are of $2+1+1$ type and therefore are fully
characterized by the pair  of particles in the composite cluster.
Obviously,  the pairs $i$, $j$ and $k$ must be
 internal to the respective two-cluster partitions, i.e.,
$i \subset \rho$, \, $j,k \subset \gamma$ and $j \subset \sigma$.
Furthermore, $\bar{\delta}_{\sigma \rho} = 1 - {\delta}_{\sigma \rho}$ and 
\begin{equation} \label{eq:G0}
G_0 = (E+i0-H_0)^{-1}
\end{equation}
 is the free resolvent at the system energy $E$.
The two-particle interactions for each pair are 
summed up to all orders to form the respective two-particle transition matrices
\begin{equation} \label{eq:t2b}
t_j = v_j + v_j G_0 t_j,
\end{equation}
while all the interactions within each two-cluster subsystem
lead to the transition operators
\begin{gather}
U_{\gamma}^{jk} = G^{-1}_0 \, \bar{\delta}_{jk} + \sum_i \bar{\delta}_{ji}
\, t_i \, G_0 \, U_{\gamma}^{ik}.
\end{gather}

The  wave function $| \Psi_{\rho,n} \rangle$ describing 
the scattering process initiated by the asymptotic
two-cluster channel state $| \Phi_{\rho,n} \rangle$
with the relative two-cluster momentum $\mathbf{p}_{\rho,n}$
is determined by the AGS transition operators as
\begin{equation} \label{eq:AGSpsi}
| \Psi_{\rho,n} \rangle = | \Phi_{\rho,n} \rangle 
  + \sum_{\gamma lki} 
G_0 \, t_l \, G_0 \,  U_{\gamma}^{lk} G_0\, t_k\, G_0 \, 
\mcu_{\gamma \rho}^{ki} | \phi_{\rho,n}^i \rangle.
\end{equation}
The energy parameter in the operators of Eq.~(\ref{eq:AGSpsi})
is $E =  \epsilon_{\rho,n} + p_{\rho,n}^2/2\mu_\rho $ 
with  $\epsilon_{\rho,n}$ being the energy of the $n$th bound state
in the partition $\rho$ and $\mu_{\rho}$ being the reduced mass.
The asymptotic channel state $| \Phi_{\rho,n} \rangle$
is a product of the $n$th bound state  wave function 
in the partition $\rho$  and the plane 
wave with momentum  $\mathbf{p}_{\rho,n}$ between the two clusters.
$| \Phi_{\rho,n} \rangle =  \sum_{i} | \phi_{\rho,n}^i \rangle$
is decomposed into its Faddeev components
that are calculated from the Faddeev equation
\begin{equation} \label{eq:phi}
| \phi_{\rho,n}^i \rangle = G_0 \sum_{j} \bar{\delta}_{ij} t_j 
| \phi_{\rho,n}^j \rangle
\end{equation}
and normalized such that
$\langle \Phi_{\sigma,n'} | \Phi_{\rho,n} \rangle = \delta_{\sigma \rho} \,
\delta_{n'n} \, \delta(\mathbf{p}_{\sigma,n'}-\mathbf{p}_{\rho,n})$.

The scattering amplitudes for the reactions with initial and
final two-cluster states are given by the 
 on-shell matrix elements of the AGS operators
$ \sum_{ji} \langle \phi_{\sigma,n'}^j |  \mcu_{\sigma \rho}^{ji} 
| \phi_{\rho,n}^i \rangle $
as explained in Refs.~\cite{grassberger:67,fonseca:87}.
The amplitude for the three-cluster breakup 
of the initial two-cluster state can be obtained from
$ \mcu_{\sigma \rho}^{ji} | \phi_{\rho,n}^i \rangle $ as well.
We derive it from the general relation
\begin{equation} \label{eq:Tj}
 \langle \Phi_{j} |  T_{j \rho} | \Phi_{\rho,n} \rangle 
= \langle \Phi_{j} | \sum_{i=1}^{6} \bar{\delta}_{ji} v_i | \Psi_{\rho,n} \rangle .
\end{equation}
The asymptotic three-cluster 
state $|\Phi_{j} \rangle $ is an eigenstate of the
channel Hamiltonian $H_0 + v_j$ with the eigenvalue $E$, i.e.,
\begin{equation} \label{eq:phij}
|\Phi_{j} \rangle  = G_0 v_j |\Phi_{j} \rangle .
\end{equation}
It is given by the bound state
wave function for the pair $j$ times two plane waves 
corresponding to the relative motion of the three free clusters.
Since the full wave function (\ref{eq:AGSpsi}) is an eigenstate of the 
Hamiltonian  (\ref{eq:H}), i.e.,
$ \sum_{i=1}^{6} v_i | \Psi_{\rho,n} \rangle = G_0^{-1} | \Psi_{\rho,n} \rangle$,
we rewrite the amplitude (\ref{eq:Tj}) as
\begin{equation} \label{eq:Tjb}
 \langle \Phi_{j} |  T_{j \rho} | \Phi_{\rho,n} \rangle 
= \langle \Phi_{j} | (1- v_j G_0) G_0^{-1} | \Psi_{\rho,n} \rangle .
\end{equation}
Although $\langle \Phi_{j} | (1- v_j G_0) = 0$ due to Eq.~(\ref{eq:phij}),
one has to keep in mind that $t_l$ arising from the wave 
function (\ref{eq:AGSpsi})
has a pole for $|\Phi_{j} \rangle $ if $l=j$. Using
 Eq.~(\ref{eq:t2b}) one gets the relation
\begin{equation} \label{eq:Tjc}
 \langle \Phi_{j} | (1- v_j G_0) t_l G_0 = \langle \Phi_{j} | \, \delta_{jl}
\end{equation}
that leads to the three-cluster breakup amplitude
\begin{equation} \label{eq:AGS0}
 \langle \Phi_{j} |  T_{j \rho} | \Phi_{\rho,n} \rangle
=  \sum_{\gamma ki} \langle \Phi_j |
U_{\gamma}^{jk} G_0\, t_k\, G_0 \, 
\mcu_{\gamma \rho}^{ki} | \phi_{\rho,n}^i \rangle.
%\end{split}
\end{equation}
Finally, the amplitude for the four-cluster breakup
can be found in Ref.~\cite{deltuva:12a}.

In the system of identical particles the number 
of distinct partitions and AGS operator components reduces to
two.  We choose those partitions to be ((12)3)4 and
(12)(34) and denote them in the following by $\alpha =1$ and $2$,
respectively. The  amplitudes (\ref{eq:AGS0})
after symmetrization become
\begin{equation} \label{eq:U0}
\begin{split}  
 \langle \Phi_{d} |  T_{d \alpha} | \Phi_{\alpha,n} \rangle 
= {}& S_{d\alpha}  \langle \Phi_{d} | 
[(1+ \varepsilon P_{34}) U_1 G_0 \, t \, G_0 \, \mcu_{1\alpha} + 
U_2 G_0 \,  t \, G_0 \, \mcu_{2\alpha} ]
| \phi_{\alpha,n} \rangle .
\end{split}
\end{equation}
Thus, $T_{d 1}$ ($T_{d 2}$) describes the atom-trimer (dimer-dimer)
breakup into a dimer and two atoms.
$S_{d1} = \sqrt{3}$ and $S_{d2} = 2$ are the 
corresponding symmetrization factors,
$P_{34}$ is the permutation operator of particles 3 and 4, 
and $\varepsilon = 1 $ (-1) for bosons (fermions). The 
dimer-atom-atom three-cluster channel state
$|\Phi_{d}\rangle $ is symmetrized only within the bound pair (12).
There is only one independent Faddeev component $| \phi_{\alpha,n} \rangle$ 
for each two-cluster channel state.
The two-particle transition matrix $t$ is derived from the pair (12)
potential according to Eq.~(\ref{eq:t2b}).
 The symmetrized $3+1$ and $2+2$ subsystem operators are
\begin{equation} \label{eq:U3}
U_{\alpha} =  P_\alpha G_0^{-1} + P_\alpha  t G_0  U_{\alpha}
\end{equation}
with $P_1 =  P_{12}\, P_{23} + P_{13}\, P_{23}$ and
$P_2 =  P_{13}\, P_{24} $ where $P_{ab}$ is the
permutation operator of particles $a$ and $b$.
The symmetrized four-particle transition operators obey the
symmetrized AGS equations \cite{deltuva:ef}, i.e.,
\begin{subequations} \label{eq:U}
\begin{align}  
\mcu_{11}  = {}&  \varepsilon P_{34} (G_0  t  G_0)^{-1}  
 + \varepsilon P_{34}  U_1 G_0  t G_0  \mcu_{11} + U_2 G_0  t G_0  \mcu_{21} , 
\label{eq:U11} \\  
\mcu_{21}  = {}&  (1 + \varepsilon P_{34}) (G_0  t  G_0)^{-1}  
+ (1 + \varepsilon P_{34}) U_1 G_0  t  G_0  \mcu_{11} , \label{eq:U21} \\
\mcu_{12}  = {}&  (G_0  t  G_0)^{-1}  
 + \varepsilon P_{34}  U_1 G_0  t G_0  \mcu_{12} + U_2 G_0  t G_0  \mcu_{22} , 
\label{eq:U12} \\  
\mcu_{22}  = {}& (1 + \varepsilon P_{34}) U_1 G_0  t  G_0  \mcu_{12} . 
\label{eq:U22}
\end{align}
\end{subequations}
 The employed basis states have to be symmetric (antisymmetric)
under exchange of two bosons (fermions) in subsystem (12) for the 
$3+1$ partition and in (12) and (34) for the $2+2$ partition.

After partial-wave decomposition 
for each combination of the total angular momentum $\mathcal{J}$
and parity $\Pi$ the AGS equations \eqref{eq:U}
become a  system of coupled   three-variable integral equations.
In our  momentum-space  framework these variables are the 
magnitudes of the Jacobi momenta $k_x$, $k_y$ and $k_z$
that are defined in  Ref.~\cite{deltuva:12a} and describe
 the relative motion in the 1+1, 2+1, and 3+1 (1+1, 1+1, and 2+2)
subsystems of the 3+1 (2+2) configurations, respectively.
Such three-variable integral
equations have been solved in  Refs.~\cite{deltuva:07a,deltuva:07c}
for various four-nucleon reactions with two-cluster initial and final states.
However,
by using a separable two-particle potential the AGS equations  \eqref{eq:U}
can be reduced to a system of integral equations in two continuous
variables, the Jacobi momenta $k_y$ and $k_z$.
We applied this technical simplification  in the study of the universal
properties of the four-boson system  that are independent 
of the short-range interaction details
\cite{deltuva:10c,deltuva:11b,deltuva:12a}. 
The same separable potential is used also in the present work.
The discretization of integrals in the AGS equations 
using Gaussian quadrature rules \cite{press:89a}
leads  to a system of linear algebraic equations.
While the calculational technique to a large extent is taken
over from Refs.~\cite{deltuva:10c,deltuva:11b,deltuva:12a},
there is one important difference associated with the
open dimer-atom-atom channel.
 Namely, in addition to the subsystem bound state poles of $U_\alpha$
the kernel of the AGS equations has integrable singularities arising from 
$t$. For $k_y^2/2\mu_{\alpha y} + k_z^2/2\mu_\alpha \to  E  - \epsilon_{d}$
the two-boson transition matrix in the channel with the dimer
quantum numbers for the pair (12) has the pole 
\begin{gather} \label{eq:tpole}
 t \to  \frac{v |\Phi_d \rangle \langle \Phi_d | v }
{E + i0 - \epsilon_{d} - k_y^2/2\mu_{\alpha y} - k_z^2/2\mu_\alpha},
\end{gather}
where $\epsilon_{d} < 0$ is the dimer energy 
and $\mu_{\alpha y}$ is the reduced mass for $k_y$.
The treatment of this singularity simplifies considerably
if the available four-boson
energy is just at the dimer-atom-atom threshold,
i.e., $E = \epsilon_{d}$. In this case there is only one
singular point $k_y = k_z = 0$ and in all nontrivial 
momentum integrals (the ones without $\delta$-functions)
of the type
\begin{equation} \label{eq:g0}
\int_0^\infty \frac{f(k_y,k_z) k_j^2 dk_j}
{i0- k_y^2/2\mu_{\alpha y} - k_z^2/2\mu_\alpha}
\end{equation}
this singularity  is cancelled by $k_j^2$ leading
to numerically harmless integrals.

\section{Dimer-atom-atom recombination \label{sec:res}}

We consider an ultracold gas consisting of a mixture of
bosonic atoms and dimers with respective 
particle densities $\rho_a$ and $\rho_d$.
We define the  dimer-atom-atom recombination
rate $K_{211}$ such that the number 
of the dimer-atom-atom recombination events 
per volume and time is $K_{211} \rho_a^2 \rho_d/2!$.  
At sufficiently low temperature  the momenta of the initial particles
are much smaller than the final momenta of the two resulting clusters, i.e.,
$k_y,k_z << p_{\alpha, n'}$, and the kinetic energies of initial particles are
much smaller than the final two-cluster  kinetic or binding
energies, i.e., $k_y^2/2\mu_{\alpha y} + k_z^2/2\mu_\alpha
<< p_{\alpha,n}^2/2\mu_\alpha, \; |\epsilon_{\alpha,n}| $.
Under these conditions one can assume the threshold values
$k_y = k_z = 0$
for the initial three-cluster channel state $ |\Phi_d \rangle$
and  $E = \epsilon_{d}$ for the AGS equations.
The dimer-atom-atom recombination rate in this limit is given by
\begin{equation} \label{eq:k211}
K_{211} = 2(2\pi)^8 \sum_{\alpha,n'}  \mu_{\alpha}  p_{\alpha, n'}
| \langle \Phi_{d} |  T_{d \alpha} | \Phi_{\alpha,n'} \rangle |^2.
\end{equation}
Here we used the time-reversal symmetry
in replacing the amplitude for the three-cluster 
recombination into the final two-cluster state
by the amplitude for the three-cluster breakup
of the initial two-cluster state, i.e.,
\begin{equation} \label{eq:t-rev}
  \langle \Phi_{\alpha,n} |  T_{\alpha d} | \Phi_{d} \rangle = 
 \langle \Phi_{d} |  T_{d \alpha} | \Phi_{\alpha,n} \rangle.
\end{equation}
Furthermore, at $k_y = k_z = 0$ only the $\mathcal{J} = 0$ partial
wave contributes to the amplitude (\ref{eq:t-rev}), i.e.,
$\langle \Phi_{d} |  T_{d \alpha} | \Phi_{\alpha,n} \rangle =
(4\pi)^{-3/2} \langle \Phi_{d} (\mathcal{J} = 0)|  T_{d \alpha} |
 \Phi_{\alpha,n} (\mathcal{J} = 0) \rangle$.

The dimer-atom-atom recombination process leads  to the  
final state either with two dimers or with an atom and an Efimov trimer.
As it is evident from Eq.~(\ref{eq:k211}),
$K_{211}$ can be decomposed into the dimer-dimer and $(n+1)$ atom-trimer
contributions
\begin{equation} \label{eq:k211x}
K_{211} = K_{211}^{dd} +  \sum_{n'=0}^n   K_{211}^{n'},
\end{equation}
where $n$ labels the highest excited trimer state.

We aim to determine the universal behaviour of the ultracold
dimer-atom-atom recombination rate and its contributions
(\ref{eq:k211x}) as functions of the two-boson scattering length.
We therefore build dimensionless quantities 
$a/a_n^d$ and $ K_{211}^x m /\hbar a^4$, with $m$ being the boson mass.
Due to asymptotic discrete scaling symmetry it is 
sufficient to consider the regime
$1 < a/a_n^d < a_{n+1}^d/a_n^d \approx 22.694$ \cite{braaten:rev}
with $n+1$ trimer states,
provided $n$ is sufficiently large 
such that non-universal short-range
corrections become negligible. We found that $n \ge 3$,
i.e., one has to consider the processes involving four or 
more atom-trimer channels.
This is fully consistent with previous findings 
\cite{schmidt:10a,deltuva:10c,deltuva:12a}
for other four-boson reactions.

\begin{figure}[!]
\includegraphics[scale=0.8]{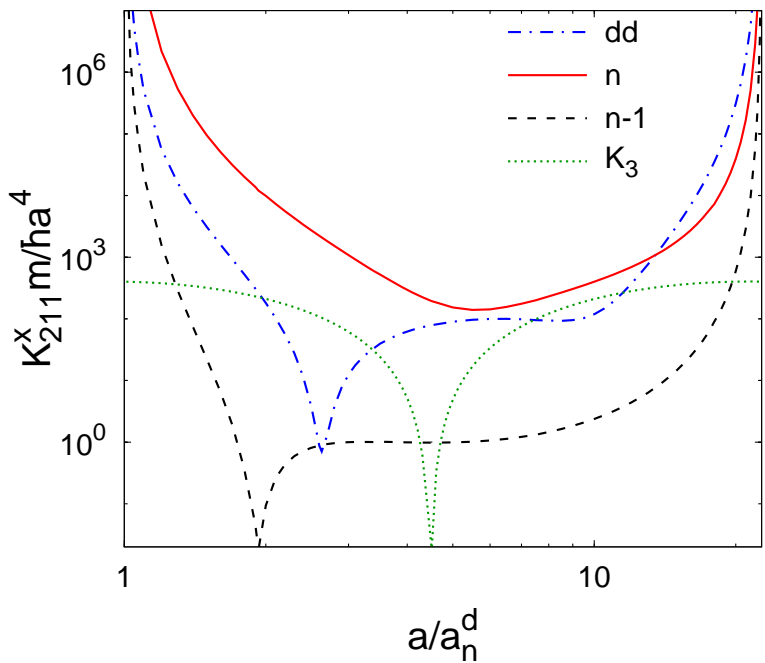}
\caption{\label{fig:k211} (Color online)
Dimer-atom-atom recombination rate 
as a function of the two-boson scattering length.
The contributions of the final dimer-dimer (dashed-dotted),
$n$-th (solid) and $(n-1)$-th (dashed) atom-trimer  channels
are compared with  the three-atom recombination rate (dotted).}
\end{figure}

We compare the most important contributions to the ultracold
dimer-atom-atom recombination rate in Fig.~\ref{fig:k211}.
All contributions peak at $a/a_n^d = 1$ and 
$ a/a_n^d \approx 22.694$ where the $n$-th and the $(n+1)$-th
 trimers, respectively,  emerge at the atom-dimer threshold.
Near these critical regimes the recombination process is dominated
by the two-cluster channel with the weakest binding,
i.e., the $n$-th atom-trimer channel
close to $a/a_n^d = 1$ and the dimer-dimer channel close to
$ a/a_n^d \approx 22.694$ with the respective
contributions  $ K_{211}^n$ and   $ K_{211}^{dd}$.
The $n$-th atom-trimer and dimer-dimer thresholds intersect
at $ a/a_n^d \approx 6.789$ \cite{deltuva:11b}. However,
it seems that there is an
interference between these two channels around
$a/a_n^d = 10$ leading to a slight enhancement of $ K_{211}^n$
and suppression of $ K_{211}^{dd}$ such that 
$ K_{211}^n > K_{211}^{dd}$ for  $ a/a_n^d < 13.5$.
This behaviour is not very surprising given the strong coupling between 
the $n$-th atom-trimer and dimer-dimer channels
 observed in the dimer-dimer scattering \cite{deltuva:11b}.
The transitions to channels
with lower trimers ($n' \le n-1$) are strongly suppressed; we only show the
predictions for  $ K_{211}^{n-1}$, the others being considerably smaller.

In addition to various $ K_{211}$ contributions we show in Fig.~\ref{fig:k211}
also the ultracold three-atom recombination rate $K_3$ defined such that
the number of three-atom recombination events per volume and time
is $K_3 \rho_a^3/3!$. We calculate it numerically as
\begin{equation} \label{eq:k3}
K_{3} = 2(2\pi)^8   \mu_{1 y}  p_{y}
| \langle \phi_{0} | (1+P_1)tG_0 U_1 | \phi_{d} \rangle |^2
\end{equation}
where $| \phi_{d} \rangle$ and $| \phi_{0} \rangle$ are the atom-dimer
and three-atom channel states and $p_y$ is the relative atom-dimer
momentum in the final state. Our $K_3$ predictions are in perfect
agreement with the semi-analytical results of Ref.~\cite{braaten:rev}
(after taking into account that the respective definitions differ
by a factor $3!$). We observe that $ K_{211}$ 
is much larger than $K_3$, especially near the  ends of the shown interval
and around the $K_3$ minimum.
This may have important consequences for the life time and stability 
of trapped ultracold atomic gases with small admixture of dimers:
Let's suppose that the trapping potential is lower than the kinetic energies
of the recombination products, i.e., they are able to escape the trap.
Neglecting other reactions like the four-atom or dimer-dimer-atom
recombination that should be suppressed at not too high densities,
the atom density in the trap evolves with time according to
\begin{equation} \label{eq:rhot}
\frac{d\rho_a}{dt} = - \frac12 K_3 \rho_a^3 - K_{211} \rho_a^2 \rho_d.
\end{equation}
Thus, due to $ K_{211} >> K_3$ even at low dimer densities $\rho_d << \rho_a$
the dimer-atom-atom recombination may be as important as the 
three-atom recombination.

\begin{figure}[!]
\includegraphics[scale=0.7]{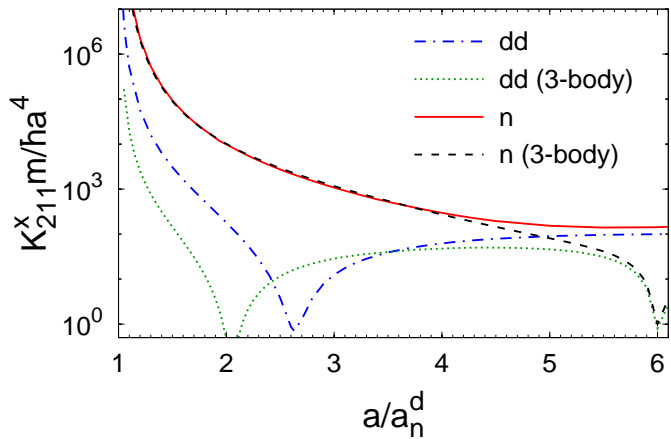}
\caption{\label{fig:k3b} (Color online)
Dimer-dimer and atom-trimer contributions
to the dimer-atom-atom recombination rate calculated using 
full four-body and approximate three-body models.}
\end{figure}

Finally, one may raise the question whether a complicated 
four-body treatment of the dimer-atom-atom recombination
is really necessary and to what extent a simple three-body model
considering the dimer as a point-like particle is reliable.
Such  model was constructed in Ref.~\cite{deltuva:12b}.
Although it describes well the $n$th trimer binding and
the low-energy atom-dimer scattering, it does not support
lower atom-trimer channels, i.e., for $n'<n$ it yields
$K_{211}^{n'} = 0$. Furthermore, the three-body dimer-atom-atom model
 is obviously inappropriate in
the dimer-dimer channel since it treats the two 
dimers asymmetrically. 
Consistently with two-cluster reactions from Ref.~\cite{deltuva:12b},
the three-body dimer-atom-atom model
fails heavily for $K_{211}^{dd}$ as shown in Fig.~\ref{fig:k3b}.
However,  $K_{211}^{n}$  in the regime
$a/a_n^d < 4$ is reproduced quite well, with 10\% or better accuracy.
There the trimer size   exceeds the  dimer size such that the 
approximations of a point-like
dimer  and a trimer consisting of weakly bound atom and dimer 
are reasonable. For  $a/a_n^d> 4$ the three-body dimer-atom-atom model
fails also in predicting $K_{211}^{n}$.

\section{Summary \label{sec:sum}}

We studied the dimer-atom-atom recombination in the universal
four-boson system that exhibits the Efimov effect.
We used exact AGS equations for the four-particle transition
operators that were solved in the momentum-space  framework.
The relation between the AGS operators and the 
dimer-atom-atom recombination amplitude was derived.
We calculated the dimer-dimer and atom-trimer contributions to the
dimer-atom-atom recombination rate in the ultracold limit.
We found that usually the channel with the weakest binding dominates
the recombination process but there are deviations from this
behaviour when several two-cluster thresholds are close to each other.
We also have shown that the dimer-atom-atom recombination rate
greatly exceeds the three-atom recombination rate leading
to important consequences for the life time of atomic gases
with a small admixture of dimers.
We also studied a simplified three-body model for the
 dimer-atom-atom recombination process and found it to be inadequate,
except for $K_{211}^{n}$ in the regime where the trimer can be
considered as a weakly bound atom-dimer system.

%%%%%%%%%%%%%%%%%%%%%%%%%%%%%%%%%%%%%%%%%%%%%%%%%%%%%%%%%%%%%%%%%%%%%%%%%%%%%
%\bibliographystyle{prsty4}
%\bibliographystyle{fewbody}
%\bibliography{abbrev,pre80,80-89,90-99,200x,clmb,ad,4N,atomic,numerics}

\end{document}